\documentclass{kapproc} 

\usepackage{procps} 
\usepackage[dvips]{graphicx}
\usepackage{epsfig}

\newcommand{\chandra}{{\it Chandra}}

\def\la{\ifmmode\stackrel{<}{_{\sim}}\else$\stackrel{<}{_{\sim}}$\fi}
\def\ga{\ifmmode\stackrel{>}{_{\sim}}\else$\stackrel{>}{_{\sim}}$\fi}
\def\farcm{\hbox{$.\mkern-4mu^\prime$}}

\begin{document}

\articletitle{Young Neutron Stars\\ and Their Wind Nebulae}

\author{Patrick Slane}
\affil{Harvard-Smithsonian Center for Astrophysics}
\email{slane@cfa.harvard.edu}

\begin{abstract}
With Teragauss magnetic fields, surface gravity sufficiently strong to
significantly modify light paths, central densities higher than that of a
standard nucleus, and rotation periods of only hundredths of a second,
young neutron stars are sites of some of the most extreme physical conditions 
known in the Universe. They generate magnetic winds with particles that are 
accelerated to energies in excess of a TeV. These winds form 
synchrotron-emitting bubbles as the particle stream is eventually decelerated 
to match the general expansion caused by the explosion that formed the 
neutron stars. The structure of these pulsar wind nebulae allow us to infer 
properties of the winds and the pulsating neutron stars themselves. The 
surfaces of the the stars radiate energy from the rapidly cooling interiors 
where the physical structure is basically unknown because of our imprecise 
knowledge of the strong interaction at ultrahigh densities. Here
I present a summary of recent measurements that allow us to infer the 
birth properties of neutron stars and to probe the nature of their winds, 
the physics of their atmospheres, and the structure of their interiors.

\end{abstract}


\section*{Introduction}

Young neutron stars (NSs) probe some of the most extreme physical environments 
in the Universe. Their rapid rotations and large magnetic fields combine
to accelerate particles to extremely high energies, producing energetic
winds that result in the slow spin-down of the stars and generate nebulae 
of synchrotron-emitting particles spiraling in a wound-up magnetic field.
The structure of these nebulae is determined by the energy input from
the central pulsars as well as the structure and content of the medium
into which they expand. In the centermost regions, relativistic outflows
in the form of rings and jets are formed; the geometry of these emission
regions reveals the orientation of the pulsar spin axes and can provide
information on the formation of kicks imparted in the moments following
their formation. Their large-scale structures reveal details of the magnetic
field and signatures of interaction with the ejecta from the explosions
that gave them birth.

The stellar interiors are characterized by conditions and physical processes 
otherwise observed only within the nuclei of atoms. They are born hot, but cool
rapidly due to neutrino production in their interiors. However, details of 
the interior structure of such stars remain poorly understood owing to 
our incomplete understanding of the strong interaction at ultrahigh densities 
and, since the neutrino production rate is critically dependent on the 
structure of the interior, the cooling rate is highly uncertain.
In the standard cooling scenario,
neutrino production proceeds primarily via the modified Urca process.
Residual heat diffuses from the core to the surface, manifesting itself as
blackbody-like emission -- modified by effects of any residual atmosphere --
which peaks in the soft X-ray band. The rate at which the surface temperature
declines depends critically upon the neutrino emission rate; thus, its
measurement provides constraints on hadronic physics at high densities.

Perhaps the most stunning thing about NSs is the fact that we
can actually make measurements that, directly or indirectly, probe the
above properties. Particularly with the advent of sensitive high-resolution
X-ray observations, we can now image jets and outflows from the wind 
termination shocks, identify magnetic filaments in the nebular interiors,
detect the thermal emission from shock-heated ejecta, and measure directly
the pulsations from the rotating stars and the emission from their ultra-hot
surfaces. Here I describe the basic properties of young NSs and
their nebulae, and summarize recent observational work that has begun a
revolution in our understanding of how these stars work. Brevity precludes
a thorough review, and the reader is referred to recent articles by
\cite{krh04} and \cite{yp04} for additional information and references.

\begin{figure}[tb]
\centerline{\epsfig{file=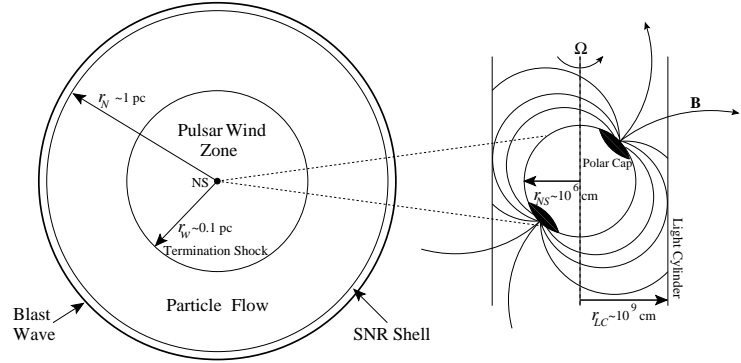,width=3.8in}}
\caption{
Schematic view of a pulsar and its wind nebula. See the text for
a complete description. (Note the logarithmic
size scaling in the PWN figure when comparing with images shown elsewhere in
the text.)
}
\end{figure}

\section{Pulsar Wind Nebulae}

Our basic understanding of PWNe stems from the picture presented
by \cite{rg74}, and expanded upon by \cite{kc84},
in which an energetic axisymmetric wind is injected from a pulsar into
its surroundings. As illustrated schematically in Figure 1, the
structure of a PWN is regulated by the input power from the pulsar and the
density of the medium into which the nebula expands; the pulsar wind and
wound-up toroidal magnetic field inflates a bubble which is confined in the
outer regions by the expanding shell of ejecta or interstellar material
swept up by the SNR blast wave. The boundary condition established by the
expansion at the nebula radius $r_N$ results in the formation of a wind
termination shock at which the highly relativistic pulsar wind is decelerated
to merge with the particle flow in the nebula. The shock forms at the radius
$r_w$ at which the ram pressure of the wind is balanced by the internal
pressure of the PWN:
\begin{equation}
r_w = \sqrt{\dot E/(4 \pi \eta c p)},
\end{equation}
where $\dot E$ is the rate at which the pulsar injects energy into the wind,
$\eta$ is the fraction of a spherical surface covered by the wind, and $p$ is
the total pressure outside the shock.  Ultimately, the pressure in the nebula
is believed to reach the equipartition value; a reasonable pressure estimate
can be obtained by integrating the radio spectrum of the nebula, using
standard synchrotron emission expressions, and assuming equipartition between
particles and the magnetic field. Typical values yield termination shock
radii of order 0.1~pc, which yields an angular size of several arcsec at
distances of a few kpc.

As the relativistic fluid comprising the PWN encounters the freely-expanding
ejecta, Rayleigh-Taylor instabilities result in the formation of a network
of dense, optical line-emitting filaments
(\cite{jun98}).
The
density and magnetic field strength becomes enhanced in regions where the
PWN encounters these filaments, producing enhanced synchrotron emission
observed
as radio filaments.
Due to the pinching effect of the global toroidal magnetic field, the overall
morphology of a young PWN is often elongated along the pulsar spin axis
(Begelman \& Li, 1992\nocite{bli92}; van der Swaluw et al.,
2004\nocite{vdk04}).
Along the rotation axis the flow becomes collimated, producing jets.
Pinch instabilities may disrupt the toroidal structure, however, changing the
structure of the magnetic field in the outer nebula regions and relaxing the
collimation of the jets far from the pulsar (\cite{beg98}).

The overall geometry of the PWN, as well as that of the emission from jets
or ring-like structures near the termination shock, thus provides a direct
indication of the pulsar geometry. The details of the jet morphology and the
emission structure in the postshock region provide the strongest constraints
available on wind composition and particle acceleration in PWNe.
For cases in which the pulsar proper
motion is also known, constraints on the kick velocity mechanism can be
derived based on the degree of alignment between the velocity vector and
the pulsar spin axis.
In later stages the PWN interacts with the reverse shock formed in the SNR
in which the NS was born. This interaction causes the disruption of the PWN,
often leading to composite SNRs with complicated PWN structures in their
interiors.

\subsection{Jets and Tori}

\begin{figure}
\centerline{\epsfig{file=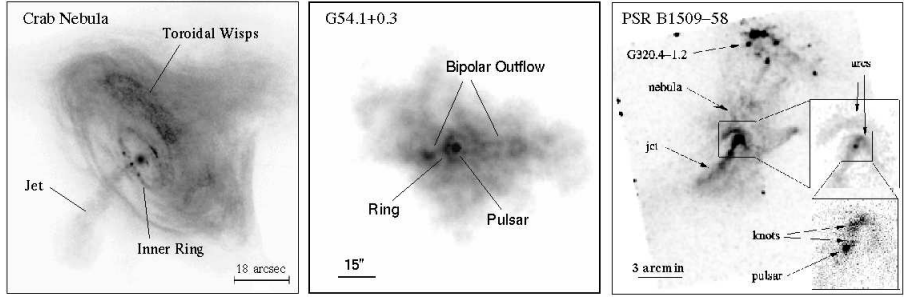,width=4.7in}}
\caption{
{\em Chandra} images of the Crab Nebula (left), G54.1+0.3 (center), and
PSR~B1509$-$58 (right) showing the complex emission
from these PWNe, including jet outflows and toroidal structures.
}
\end{figure}

In the inner portions of the Crab Nebula, optical wisps mark the position of
the wind termination shock, at a distance of $\sim 0.1$~pc from the pulsar.
The brightness and position of these wisps varies
in time, with inferred outflow speeds up to $0.7 c$
(\cite{hes98}).
As shown in Figure 2 (left),
high resolution X-ray images reveal a ring of emission at
the position of the wisps (\cite{wht+00}),
providing a direct
connection between the unshocked pulsar wind and the bulk properties of
the nebula. Material from the inner ring forms a series of toroidal X-ray
wisps that are variable with time (\cite{hmb+02}).
The geometry of these
X-ray features imply a tilted torus, and a jet of material flows perpendicular
to the plane of the toroid, extending some 0.25~pc from the pulsar. A faint
counterjet is also observed, along with significantly enhanced X-ray
emission from the leading portion of the toroid, presumably the result
of Doppler beaming. 
One troubling aspect of this suggestion is that the
brightness distribution around the inner ring does not match that of the
outer toroid; indeed, the brightness is rather uniform except for some
small clump-like structures that vary in position and brightness with time.

A handful of other PWNe display X-ray features that suggest the presence of
extended ring-like structures and narrow collimated components.
The size of the ring-like features places the emission region near the
pulsar wind termination shock. The spectral and temporal properties of
the collimated structures argue that they are focused jets of high speed
material, as observed in the Crab. Such observations have already begun
to inspire new axisymmetric MHD models that predict similar features
(e.g., Komissarov \& Lyubarsky 2004\nocite{kl04}),
and ongoing observational studies
promise to further constrain and refine such models. In particular, the
confining mechanism for jets is not well-understood; many jets display some
amount of curvature, with the Vela pulsar jet being an extreme example
in which the morphology is observed to change on timescales of months
(\cite{ptk+03}).
This may be the result of pinch instabilities disrupting the toroidal
structure of the confining magnetic field (Begelman 1998), or could be
indicative of an interaction of the jet material with the ambient medium.
There also appears to be a wide variation in the
fraction of spin-down energy channeled into the jets, ranging from roughly
$2.5\times 10^{-5}$ for
PSR~J0205+6449 in 3C~58 to nearly $10^{-3}$ for PSR~B1509$-$58. And,
while Doppler
beaming is invoked to explain the large brightness variations in jets and
the associated counterjets, as well as around the observed toroidal
structures, it is not clear that this alone is sufficient to explain the
observations. 

\chandra\ observations of G54.1+0.3 (\cite{lwa+02}) reveal
a central 136~ms pulsar (\cite{clb+02}) embedded
in a diffuse
$1\farcm5 \times 1\farcm2$ nebula (Figure 2, center).
The pulsar is surrounded by an X-ray ring for which the
X-ray emission is brightest along the eastern limb. When interpreted
as the result of Doppler boosting, this implies a post-shock velocity
of $\sim 0.6c$ (Lu et al. 2002; Romani \& Ng 2003\nocite{rn03}).
Faint bipolar elongations running roughly east-west, perpendicular to the long 
axis of the ring, are also observed. These apparent
outflows, which presumably lie along the pulsar rotation axis, are more
diffuse than the jets in the Crab Nebula, yet appear to carry away a
considerably larger fraction of the energy;
they comprise roughly the same luminosity as the central ring,
which is in stark contrast to the Crab where the torus outshines the jets
by a large factor. 

\chandra\ observations of PSR~B1509$-58$ (\cite{gak+02})
demonstrate that this young and energetic
pulsar associated with G320.4--1.2 powers
an extended and extremely complicated PWN, with
structures on scales from $\sim10'$ down to the spatial
resolution limit (Figure 2, right).
The elongated PWN has a clear axis
of symmetry centered on the pulsar, presumably representing the projected
orientation of the pulsar spin axis.  To the southeast of the
pulsar, the nebula is dominated by a narrow jet-like feature
approximately 6~pc in length.
The lack of a similar feature to the north can be explained by Doppler
boosting if the pulsar's spin axis is inclined
to the line-of-sight by $\la 30^\circ$ (Gaensler et al., 2002).
In the central regions of the PWN, a pair of semi-circular arcs lie $\sim0.5$
and $\sim1$~pc to the north of the pulsar. Gaensler et al. (2002) note that
if the inner region of these arcs represents the position
of the pulsar wind termination shock, then the flow time to the arcs
is much shorter than the synchrotron lifetime of the emitting particles
based on equipartition estimates of the magnetic field. Thus, unlike for
the Crab torus, where these timescales are similar, the emission from
the arcs is not the result of large synchrotron cooling at this position.
Instead, the arcs appear to resemble the series of concentric wisps seen
for the Crab which are interpreted as sites of electron compression in an 
ion-dominated flow (\cite{ga94}, Gaensler et al., 2002).

The innermost region of 3C~58 (see Figure 3) consists of a bright, elongated
compact structure centered on the pulsar J0205+6449. This inner nebulosity
is bounded along the western edge by a radio wisp
(\cite{fm93}),
and is suggestive of a toroidal structure that is tilted about
a north-south axis, with the pulsar at its center. 
The eastern side of the toroid is slightly
brighter than the western side, suggesting that the eastern side is beamed
toward us. If interpreted as a circular termination shock
zone, the inferred inclination angle in the plane of the sky is roughly
70 degrees (\cite{shm02}).

\begin{figure}
\centerline{
\epsfig{file=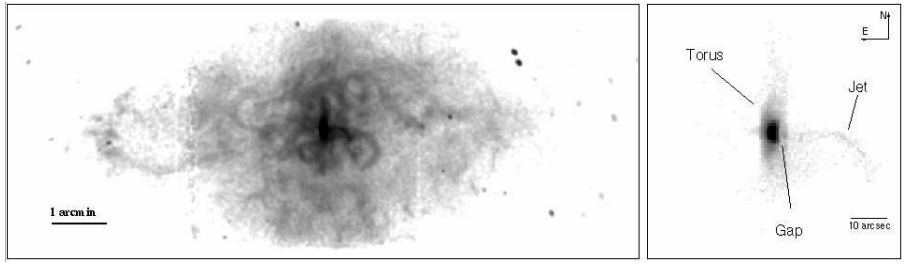,width=4.7in}}
\caption{
Left:
{\em Chandra} image of 3C 58. Complex filamentary loops fill the interior
region.  Right: The innermost region of 3C~58 showing the NS embedded in an
elongated structure. A curved jet extends to the west, with a hint of a
counterjet component in the east.
}
\end{figure}

The elongated structure extending westward from the position of the pulsar has
the appearance of a jet (Figure 3, right). Its orientation is consistent,
in projection, with the pulsar rotation axis inferred from the wind
termination shock region discussed above, and also the east-west elongation
of the entire PWN (Figure 3, left). The structure shows considerable 
curvature, similar to
that seen in the Crab Pulsar jet. 
A faint structure that may be a counterjet is observed to the
east of the pulsar. The observed luminosity is
nearly a factor of 10 smaller than that for the torus.
For the Crab Nebula, the torus is nearly 20 times more luminous than the
jet in X-rays, while for PSR~B1509--58 the jet is brighter than the extended
inner emission (Gaensler et al. 2002).

The jet/torus morphology observed in these PWNe provides the geometry of the
pulsar system, yielding both the projected direction of the spin axis and
the inclination angle. Modeling of such emission in other PWNe holds promise
for understanding the kicks that give pulsars their large space velocities
(\cite{nr04}).
The jets observed in the Crab and Vela pulsars, for example, are
aligned with their
proper motion vectors (Aschenbach \& Brinkman 1975\nocite{ab75};
Helfand et al. 2001\nocite{hgh01}).
If the kick that gave these pulsars their proper motion was
generated in the supernova explosion by some asymmetric mass ejection,
then this alignment requires an initial pulsar spin period that is
short relative to the kick timescale, so that the impulse of the kick
is averaged over many rotations of the star (\cite{lcc01}).
\cite{rn03} reach similar conclusions for PSR J0538+2817 in
the supernova remnant (SNR) S147. By modeling the faint extended PWN emission
as a jet and torus, they derive a spin axis direction that is aligned with
the vector from the SNR center to the current pulsar position.
For some pulsars [e.g. J0205+6449 in 3C~58 (\cite{mss+02})
and J1811--1925 in G11.2--0.3 (\cite{krv+01})],
we believe that the initial spin period was much longer
than typical pulsar kick timescales.
This would suggest that their proper
motions should not necessarily be aligned with the jet direction.
Future radio timing observations of these pulsars
will ultimately lead to such proper motion measurements.

\subsection{Filaments in PWNe}

Extensive filamentary structure is observed in H$\alpha$, [OIII], and other
optical line images of the Crab Nebula. Based on their observed velocities,
these filaments form an expanding shell of ejecta that surrounds the
nonthermal optical emission from the nebula. High resolution
images with HST reveal detailed morphology and ionization structure
suggesting that the filaments form from Rayleigh-Taylor instabilities
as the expanding relativistic bubble encounters slower moving ejecta
(\cite{hss96b}), a picture supported by MHD
simulations that show that 60-75\% of the swept-up mass ends up
concentrated in such filaments (Jun 1998\nocite{jun98},
Bucciantini et al. 2004\nocite{bab+04}).
Radio observations reveal filaments that coincide with these optical
filaments, presumably corresponding to synchrotron emission from regions
of enhanced density and magnetic field in the form of magnetic sheaths that
form as the pulsar-injected energy encounters the thermal filaments
(\cite{rey88b}).
Such filamentary structure is not observed in X-rays, however,
suggesting that the electrons with sufficient energy to radiate X-rays do not
reach the shell of filaments. This is consistent with the observed smaller
extent of the X-ray emission in the Crab nebula relative to its radio size,
and indicates a larger magnetic field than is observed in 3C~58 and
PSR~B1509$-$58.

Recent \chandra\ observations of 3C~58 reveal a complex of loop-like
filaments most prominent near the central regions of the PWN (Figure 3, left),
but evident throughout the nebula (\cite{shvm04}).
These structures,
whose X-ray spectra are nonthermal, are very well correlated with features
observed in the radio band (\cite{ra88}).
Optical observations reveal faint thermal filaments as well
(\cite{van78}), which presumably have an origin similar to that of the
Crab filaments. The velocities of these optical filaments in 3C~58
are $\sim \pm 900{\rm\ km\ s}^{-1}$
(\cite{fes83}), sufficiently
high to indicate that the PWN is young, but too small to account for the
current size of 3C~58 if the historical age is assumed -- one of
several standing problems with regard to its evolution (\cite{che04}).
A detailed comparison
of the X-ray and optical images shows that most of the X-ray filaments do not
have corresponding optical structures, however. While comparisons with
deeper optical
images are clearly needed, the fact that many of the X-ray features without
optical counterparts are brighter than average in X-rays suggests that these
may actually arise from a different mechanism. Slane et al. (2004) propose
that the bulk of the discrete structures seen in the X-ray
and radio images of 3C~58 are magnetic loops torn from the toroidal field
by kink instabilities. In the inner nebula, the loop sizes are similar to
the size of the termination shock radius, as suggested by Begelman (1998).
As the structures expand, they enlarge slightly as a consequence of the
decreasing pressure
in the nebula. Some of the observed X-ray structure in the outermost regions
may be the result of thermal filaments produced by Rayleigh-Taylor
instabilities, similar to the filaments in the Crab Nebula. A shell
of thermal X-ray emission demonstrates the presence of ejecta in these outer 
regions (\cite{bwm+01}, \cite{shvm04}. 

It is worth noting that considerable loop-like filamentary structure 
is evident in \chandra\ observations of the Crab Nebula as well
(\cite{wht+00}). These
features are primarily observed encircling the bright Crab torus, perpendicular
to the toroidal plane, and may result from currents within the torus itself.
It is at least conceivable that such currents are signatures of the kink
instabilities suggested above.

\subsection{Large-Scale Structure of PWNe}

\begin{figure}
\centerline{\epsfig{file=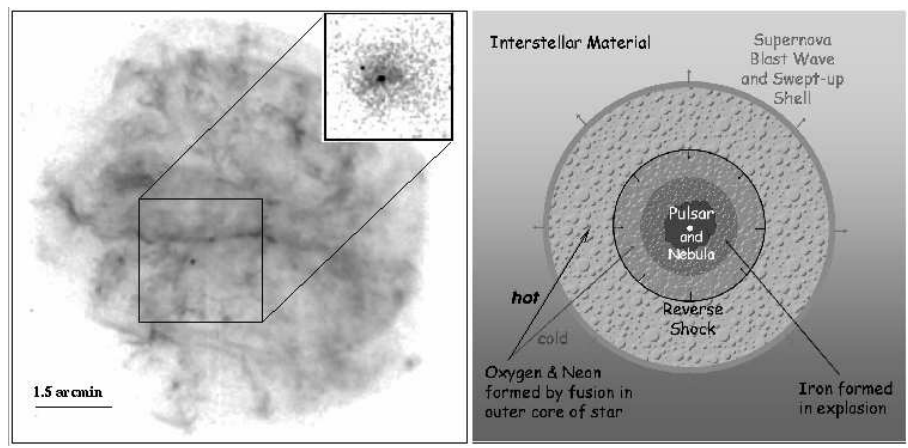,width=4.7in}}
\caption{
Left: {\em Chandra} image of G292.0+1.8. The inset shows the central region
at energies above 4~keV, where the pulsar and its wind nebula dominate.
Right: Schematic diagram of the evolutionary state of G292.0+1.8. The lack
of iron observed in the spectrum indicates that the reverse shock has not
yet made its way to the center of the remnant, where the PWN resides.
}
\end{figure}

The large-scale elongated shape of 3C~58 is similar to that found (particularly
in the radio band) for a number of other PWNe including the Crab Nebula and
G54.1+0.3. Magnetohydrodynamical calculations by Begelman \& Li, 1992,
and van der Swaluw, 2004, show that such an elongation can result from the
pinching effect of a toroidal magnetic field for which the projected axis
lies along the
long axis of the PWN. The pinching effect results in a low pressure at the
edge of the bubble along the major axis with respect to the (higher) pressure
at the edge of the minor axis, which yields the elongated structure.
The elongation thus marks the projection of the spin
axis of the pulsar producing the wound-up field.  In 3C~58 this is
consistent with the inference of an east-west direction for the projected
spin axis based on the interpretation of the extended structure in the inner
nebula as being associated with a tilted ring-shaped wind termination shock
zone (\cite{shm02}).

The structure of a PWN can be altered significantly through interaction
with the reverse shock from the SNR in which it resides. In its early 
evolution the PWN is basically freely-expanding, encountering only small
amounts of slow-moving ejecta in the SNR interior. As the SNR blast wave
sweeps up sufficient amounts of circumstellar/interstellar material, a
reverse shock is driven back through the ejecta. As this reverse shock
propagates, heating the ejecta, it will eventually reach the PWN. 
\chandra\ studies of the oxygen-rich remnant G292.0+1.8 reveal an SNR
in the intermediate stages of this process. The 0.5-10~keV X-ray image is 
presented in Figure 4 (left), and shows the complex structure associated with
the shock-heated ejecta and CSM
(\cite{prh+02}).
The inset shows the central image at energies above 4~keV, and reveals a
compact pulsar surrounded by a wind nebula (\cite{hsb+01}, \cite{cmg+02},
\cite{hsp+03}). X-ray spectra of the SNR show metal-rich ejecta with strong
lines of oxygen and neon, but a distinct shortage of iron emission
(\cite{phs+04}), indicating that the reverse shock has not yet 
propagated sufficiently far toward the center to heat the iron-rich
material that was formed closest to the core of the progenitor 
(Figure 4, right).

\begin{figure}
\centerline{
\epsfig{file=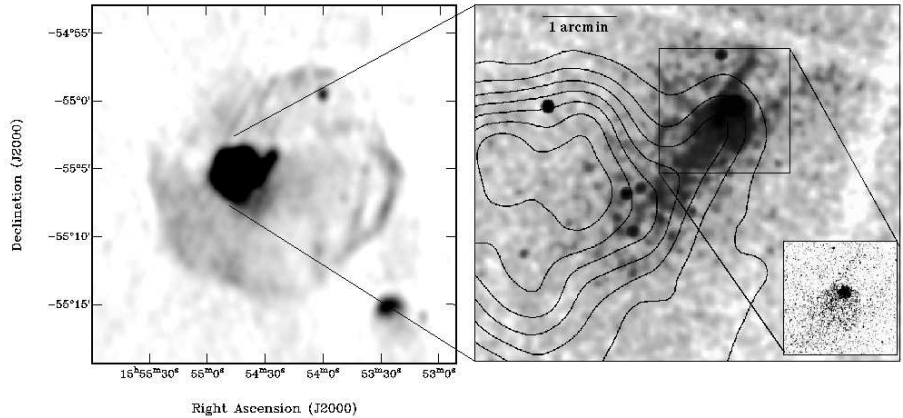,width=4.7in}}
\caption{
Left: MOST image of the composite SNR G327.1$-$1.1.
Right: {\em Chandra} image of the PWN in G327.1$-$1.1 (with radio contours),
showing diffuse emission surrounding a compact, but resolved, X-ray source
(inset).
}
\end{figure}

The morphology of the PWNe in G327.1$-$1.1 presents a rather different
picture. The radio image (Figure 5, left) reveals a well-defined SNR shell 
with a bright PWN in its interior, distinctly offset from the geometric
center. A finger of emission extends to the northwest of the radio PWN, and
\chandra\ observations reveal a compact X-ray source at this location (Figure 5,
right). The source is slightly extended (see inset) suggesting that we are 
seeing material near the wind termination shock. The compact source 
resides at the tip of a trail of emission that leads back to the bulk of the 
radio nebula, rather than at the center of the PWN (as in G292.0+1.8, for
example), suggesting that the PWN morphology results from a combination 
of the pulsar motion and the passage of the reverse shock which has
apparently disrupted the western side of the nebula (Slane et al.,
in preparation).

\section{Neutron Star Cooling}

The cooling rate of isolated NSs has been a subject of
considerable theoretical work predating even the discovery of the first
pulsars (e.g., Bahcall \& Wolf 1965\nocite{bw65}). 
The poorly understood properties of
the strong nuclear potential at the densities found in NS interiors
make these calculations difficult, and lead to a wide range of predictions
based on different assumptions for the equation of state, composition, and
details of superconductivity (see, e.g., reviews by Tsuruta,
1998\nocite{tsur98}, and Yakovlev \& Pethick, 2004\nocite{yp04}). While 
there is a clear consensus that the early cooling proceeds via neutrino 
emission from the NS core, the timescale over which this dominates depends 
critically on the neutrino production rate which, in turn, can vary by 
orders of magnitude depending upon the state of matter in the interior.

Broadly speaking, models can be divided into ``standard'' and
``non-standard'' cooling scenarios based on the rate of neutrino
production in the NS interior. 
At the high densities and low proton fractions expected in NS interiors, 
the direct Urca reactions
($n \rightarrow p + e^- + \bar\nu_e$ and $p + e^- \rightarrow n + \nu_e$)
cannot conserve both energy and momentum.
Instead, a bystander baryon is required for each interaction
to absorb momentum. The neutrino rate for this so-called modified Urca
(mUrca) process is considerably lower than in the direct (dUrca) process
because of the extra interaction required, and is the basis for standard 
cooling models.
neutrino bremsstrahlung, and plasmon neutrino processes). 
In Figure 6 (left), we plot cooling curves for different models of
the NS interior and its properties using representative neutrino rates
(see \cite{pag98} and references therein).
The solid curve corresponds to ``standard''
cooling using an equation of state of moderate stiffness.

Non-standard cooling models incorporate neutrino emissivities
associated with other processes that may operate in NS interiors, such as those
arising from the presence of pion condensates which may form
at sufficiently high densities. The resulting pion-induced
beta decay leads to very a high neutrino emissivity and a correspondingly
shorter cooling time for the NS interior.  Similar processes involving kaon
condensates or quark matter may operate as well. Alternatively, 
equations of state that allow a high proton fraction in the interior
may allow the dUrca process to proceed; this
also leads to extremely high neutrino production rates (e.g., \cite{kyg02}).
These nonstandard cooling
mechanisms modify the NS cooling curves substantially. The dashed curves
in Figure 6 represent approximations for several nonstandard cooling models
and illustrate the associated rapid cooling (Page 1998).

\begin{figure}[t]
\centerline{\epsfig{file=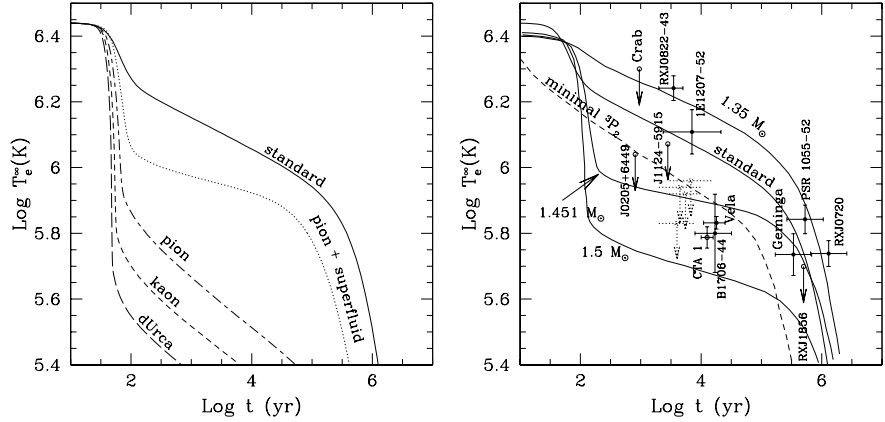,width=4.7in}}
\vspace{-0.2in}
\caption{
Left: Representative neutron star cooling curves for neutrino rates
corresponding to different interior conditions. Data from Page, 1998.
Right: Temperature vs. Age measurements for neutron stars, along with
plots for standard cooling (bold), ``minimal'' cooling, and mass-dependent
cooling invoking dUrca (see text).
}
\vspace{-0.2in}
\end{figure}

The effects of superfluidity can substantially moderate such rapid cooling
because the significantly reduced heat capacity of the superfluid particles
reduces the neutrino rate considerably. Thus, slow cooling can occur from
a combination of slow neutrino rates and a high degree of superfluidity, while
rapid cooling places strong constraints on both, and thus on the structure
and physics of the NS interior, as well as on the microphysics of superfluidity.
In particular, as the temperature approaches the critical temperature for
the superfluid state, the formation and breaking of Cooper pairs opens another
channel for neutrino emission that can lead to more rapid cooling.
Page et al., 2004\nocite{plps04} have calculated a ``minimal cooling'' scenario
which extends the standard mUrca cooling scenario to include the contributions
from this Cooper pair process. As discussed below, for some superfluidity
models they find sufficiently rapid cooling to explain most observations
of young NSs. \cite{yp04}, on the other hand, have considered both standard
and enhanced cooling models. Using the NS mass as a free parameter, they find
models for which the dUrca process becomes active for sufficiently high masses,
thus leading to a picture in which the young, cool NSs correspond to
those with higher mass (see below).

\subsection{Measuring Neutron Star Temperatures}

Treating the emission from the NS surface as a blackbody, X-ray spectral
fitting provides a measure of the gravitationally redshifted temperature
and luminosity (assuming the distance is known):
\begin{equation}
T_s^\infty = \left[1 - \sqrt{\frac{2GM}{Rc^2}}\right] T_s;
L^\infty = \left[1 - \sqrt{\frac{2GM}{Rc^2}}\right]^2 L
\end{equation}
where the quantities on the left are observed at infinity, and those on the
right are at the NS surface; $M$ and $R$ are the mass and radius of the NS.
From this we calculate the effective radius, which can be compared
directly with predications for different equations of state:
$R^2_{eff} = L^\infty/[4 \pi \sigma (T_s^\infty)^4]$
(where $\sigma$ is the Stefan-Boltzmann constant). Alternatively, if only
an upper limit on the source luminosity is determined, a temperature upper
limit can be derived by assuming a value for the NS radius.

As with all stars, the emission from the surface of a NS is not a blackbody;
rather, it is modified by the presence of whatever atmosphere might exist.
One expects the surface of the NS to be covered with Fe, but an atmosphere
consisting of H, He, and/or intermediate-mass elements acquired either from
ejecta fallback following
the neutron star's formation, or from material accreted from the ISM, is also a
possibility. From models of nonmagnetic atmospheres, the primary
effect of H or He atmospheres is a considerable deviation of the
high energy end of the spectrum relative to the Wien tail of a pure
blackbody. The result is that attempts to fit the observed
emission with a blackbody model will overestimate the effective temperature
-- typically by as much as a factor of two. The
inferred size of the NS would, in turn, be underestimated in order to
yield the same flux.  For atmospheres
dominated by heavier elements the effect is considerably reduced, and the
blackbody fit gives a good approximation to the temperature.

\subsection{Confronting Cooling Models}

In Figure 6 (right) we plot the measured temperatures, or upper limits, for
the pulsars and compact objects in SNRs for which these values are best
determined. Values for known pulsars are plotted with closed circles. For
comparison, curves are shown for standard cooling as well as the ``minimal''
cooling model of \cite{plps04} (dashed curve). The latter model assumes
no enhanced cooling mechanisms in the interior. The primary mechanism
which results in faster cooling is neutrino emission from the Cooper pair
breaking and formation process, which is heavily dependent upon the assumed
superfluidity model (here we have plotted their model ``a'' for the neutron
$^3$P$_2$ gap); the predicted cooling rate is sufficiently rapid to explain
the measurements for most pulsars, although the temperatures for the Vela
Pulsar and J0205+6449 in 3C~58 fall appear to require more rapid cooling.
The inferred temperature for RX~J0007.0+7302 (plotted as an open box in Figure
6), a compact X-ray source in the SNR CTA~1, also falls below this minimal 
cooling scenario (Slane et al., 2004\nocite{szhs+04}). While pulsations 
have yet to be detected 
from this source, \chandra\ observations by \cite{hgch+04} reveal a slightly 
extended source with a distinct jet-like tail that is the signature of the 
pulsar outflows described in Section 1.2. 

For comparison, cooling curves illustrating the effects of increasing mass
are plotted as solid curves, using the model 1p for proton superfluidity
from Yakovlev and Pethick, 2004.  Also plotted in the Figure, as dashed
arrows, are upper limits based on a neutron star search in nearby SNRs
(\cite{kfgg+04}). In this study, deep X-ray observations were used to identify,
sources within the field of the SNRs, and follow-up optical and IR 
observations were used to screen for non-NS counterparts. While the lack
of detection of a NS in these remnants (G127.1+0.5, G84.2-0.8, G93.3+6.9, and
G315.4-2.3 in increasing order of temperature upper limit) may indicate
that these SNRs originated in Type~Ia events, or all formed black holes,
this is statistically unlikely. Future measurements from this study will
solidify this picture, and perhaps provide further evidence for rapid cooling
in young neutron stars.

\section{Summary}
Due in large part to the availability of high resolution X-ray measurements,
the study of young NSs and their PWNe have yielded dramatic new information
on the nature of the stellar interiors and the structure of their winds. It
is now clear that the axisymmetric wind from a pulsar goes through a 
termination shock as it joins the slower flow of its extended nebula, and
that jets and toroidal structures characterizing the inner structure can
be used to infer the orientation of the pulsar spin axis. The brightness
variations in the inner nebula and jets, as well as the spectral and spatial
structure of these regions, yield information on the outflow geometry as well
as the fraction of spin-down energy being channeled into these regions. With
broad application to particle acceleration and jet formation in 
astrophysical settings, these observations are providing constraints on
theoretical models of considerable importance for a wide range of problems.
Similarly, new observations are providing unprecedented capabilities for
detecting young neutron stars in SNRs, and for characterizing their emission.
These have forced a revised look at models for the structure of NS interiors
and provide the best opportunity for addressing the possibility that 
exotic states of matter reside in NS cores. Through additional 
and more sensitive observations of these systems, we anticipate significant
refinements, and undoubtedly new surprises, in broad picture of their
structure that is currently unfolding.

\begin{acknowledgments}
The author wish to thank Bryan Gaensler, David Helfand, Jack Hughes, and 
Fred Seward for their particular contributions to this work. Informative
discussions on NS cooling with Dima Yakovlev and Dany Page are also gratefully
acknowledged. This work was supported in part by NASA contract NAS8-39073
and grants GO0-1117A, NAG5-9281, and GO1-2054X.
\end{acknowledgments}

\end{document}